\documentclass[%
 reprint,
 amsmath,amssymb,
 aps,
]{revtex4-2}
\usepackage{hyperref}
\usepackage{cleveref}
\usepackage{graphicx}

\begin{document}
\title{How universal are warm turbulent cascades?}
\author{Ben Y. Israeli}
\email{ben.israeli@weizmann.ac.il}
\affiliation{
Weizmann Institute of Science\\
Department of Physics of Complex Systems
}
\author{Gregory Falkovich}
\affiliation{
Weizmann Institute of Science\\
Department of Physics of Complex Systems
}
\begin{abstract}
It is often assumed that general stationary solutions to the wave kinetic equation are described solely by conserved fluxes and temperature.
The families of solutions parameterized by these quantities are referred to as warm cascades, and are derivable from differential models of wave kinetics.
These models rely on the assumption of \emph{local} interactions, and  we demonstrate by a time scale argument that the \emph{nonlocal} effect of thermal damping on the cascade may result in dependence on an additional parameter, the UV cutoff, calling into question the validity of warm cascade solutions.
This is shown for systems with both three-wave and four-wave interactions.
We then consider the depletion of the cascade by this damping, and present a numerical demonstration of this nonlocal and nonuniversal behavior.
\end{abstract}

\maketitle
\section{Introduction}
There are two classes of universal states of wave systems: thermal equilibrium defined by the temperature and turbulent cascade defined by a flux. It is often assumed that the general stationary solution of the wave kinetic equation is determined by free parameters describing conserved fluxes and temperature. Such general solutions are called warm cascades, and they are physically quite tempting since they would describe in one sweep both the cascade towards the heat bath and the modes close to equilibrium with the bath~\cite{nazarenkoWaveTurbulence2011,zakharovKolmogorovZakharovSpectraTurbulence2025}.
They are generally derived using differential models of turbulence, which assume that interactions are superlocal in $k$-space~\cite{boffetta_modeling_2009,connaughton_warm_2004,dyachenkoOpticalTurbulenceWeak1992,leith_diffusion_1968,lvov_differential_2006,lvov_gradual_2008,nazarenko_differential_2006}.

Here, we call into question the concept of warm cascades as a universal family determined by fluxes and temperature (and chemical potential when it is nonzero). 
The reason is nonlocal interaction in wavenumber space. Indeed, cascade spectra presume local interactions, whereas thermal equilibrium is achieved for all interactions. When cascade and equilibrium coexist, they generally interact nonlocally.
This results in dependence of the solution on additional parameters, and in some cases results in an intermediate region with a non-universal spectrum.
We derive conditions for such nonlocality in three- and four-wave systems, which motivates an approximate form for the impact of nonlocal damping on the energy cascade and the resulting transition.
We  then focus on acoustic turbulence in an isotropic continuous medium as a detailed example~\cite{zakharovSpectrumAcousticTurbulence1970}.
We show numerically that the behavior of the transition between the cascade and thermal ranges is dependent on the maximum wavenumber and is consistent with our model.

\subsection{The wave kinetic equation}
In its most general form, the wave kinetic equation may be written
\begin{equation}
    \partial_t n_{\boldsymbol{k}}=I_{\boldsymbol{k}}[n]
    \label{equ:WKE}
\end{equation}
where $n_{\boldsymbol{k}}$ is the wave action or occupation number for wave vector $\boldsymbol{k}$, and $I_{\boldsymbol{k}}[n]$ is a collision integral describing the coupling of mode $k$ to other modes.
For the case of three-wave interactions, the collision integral takes the form
\begin{widetext}
\begin{equation}
I_{\boldsymbol{k}}[n]= \pi \int\left[\left|V_{k 12}\right|^2 f_{k 12} \delta\left(\boldsymbol{k}-\boldsymbol{k}_1-\boldsymbol{k}_2\right) \delta\left(\omega_k-\omega_1-\omega_2\right)-2\left|V_{1 k 2}\right|^2 f_{1 k 2} \delta\left(\boldsymbol{k}_1-\boldsymbol{k}-\boldsymbol{k}_2\right) \delta\left(\omega_1-\omega_k-\omega_2\right)\right] d \boldsymbol{k}_1 d \boldsymbol{k}_2
\label{equ:I-3}
\end{equation}
\end{widetext}
where
\begin{equation}
f_{k12}\equiv n_1 n_2-n_{\boldsymbol{k}}\left(n_1+n_2\right),\ n_j=n_{\boldsymbol{k}_j},
\end{equation}
and $V$ is a function of momenta defining the wave-wave coupling.
For the four-wave case, it is
\begin{equation}
\begin{split}
     \partial_tn_k=\pi\int &d\vec k_1\,d\vec k_2\,d\vec k_3\,|V_{k123}|^2\\
     &f_{k123}\,\delta_{k1,23}\,\delta(\omega_k+\omega_1-\omega_2-\omega_3)
\label{equ:WKE_four-wave}
\end{split}
\end{equation}
where
\begin{equation}
    f_{k123}=n_2n_3\left(n_k+n_1\right)-n_kn_1\left(n_2+n_3\right)
\end{equation}

Generally, the wave kinetic equation has two categories of known stationary solutions: Kolmogorov-Zakharov (KZ) cascades of conserved fluxes (energy in the three-wave case and both energy and wave action in the four-wave case), and thermal equilibrium given by
\begin{equation}
n_{\boldsymbol{k}}=\frac{T}{\omega(\boldsymbol{k})}
\end{equation}
for waves with frequency $\omega(\boldsymbol{k})$.
For a system in $d$ dimensions with homogeneous $V$ of degree $m$, the direct energy cascades for the three- and four-wave cases are, respectively
\begin{align}
    n_k=&\frac{C}{V_0}P^{1/2}k^{-m-d},
    \label{equ:three-wave-cascade}\\
    n_k=&\left(\frac{C}{V_0}\right)^{2/3}P^{1/3}k^{-d-2m/3},
    \label{equ:four-wave-cascade}
\end{align}
with dimensionless constant $C$~\cite{zakharovKolmogorovZakharovSpectraTurbulence2025}.

\subsection{Differential models}
Warm cascades, which have  thermal and cascade asymptotics respectively at small and large wavenumbers, can be derived using differential models, which assume superlocal interactions~\cite{connaughton_warm_2004}.
For example, in the case of wave turbulence with a three-wave interaction, they may be motivated as follows.
 The wave kinetic equation may be written as a continuity equation for energy density and energy flux in $k$-space.
\begin{equation}
\partial_t \mathcal{E}(\boldsymbol{k})=-\partial_{\boldsymbol{k}} P
\label{equ:energy_continuity}
\end{equation}
For superlocal interactions, the flux is proportional to (minus) the local gradient, so that the kinetic equation turns into a diffusion equation. In the isotropic case, parameterizing by frequency, we have
\begin{equation}
\partial_t \mathcal{E}=\partial_\omega D(\omega,\mathcal{E}) \partial_\omega \mathcal{E}
\end{equation}
where for three-wave interaction the diffusion coefficient should be linear in $\mathcal{E}$ and (asymptotically) go as some power of $\omega$, $D(\omega,\mathcal{E})\sim\omega^a \mathcal{E}$. The general stationary solution corresponds to a constant flux and depends on two parameters:
\begin{equation}
    \omega^a\mathcal{E}{\partial \mathcal{E}\over\partial \omega}=-P\ \Rightarrow\ \mathcal{E}=\sqrt{T^2+2P\omega^{1-a}/(a-1)}\ .\label{diff1}
\end{equation}
Here, evidently $a>1$ when $P>0$, since the positive flux corresponds to a cascade steeper than equipartition.
Then the solution (\ref{diff1}) exactly fits the simple expectations: it is equipartition at $\omega\to0$, cascade at $\omega\to\infty$, and is universal, that is determined by the flux and temperature only.

To match the KZ spectrum $E\sim \omega^x$ for a particular system, one chooses $a=1-2x$.
We may also rewrite the differential equation in terms of the wave action as
\begin{equation}
    \partial_t n(k)=C k^{1-2\alpha} \partial_k k^{1-\alpha(1-2x)}n(k)\partial_k k^{-\alpha} n(k),
\end{equation}
where $\omega\sim k^\alpha$.

We find that, with the assumption that interactions are strongly local, the problem is reduced to a differential equation with a simple family of solutions which asymptotically match cascade and equilibrium solutions.

The differential approximation has been applied to optical and Kelvin wave turbulence~\cite{boffetta_modeling_2009,dyachenkoOpticalTurbulenceWeak1992,lvov_gradual_2008,nazarenko_differential_2006}, and for hydrodynamic turbulence in two and three dimensions~\cite{connaughton_warm_2004,leith_diffusion_1968,lvov_differential_2006,lvov_gradual_2008}, yet in each case, it is an uncontrolled approximation. That means that one cannot guarantee even qualitative features of turbulence being faithfully reproduced.
Indeed, it has been noted that for hydrodynamic turbulence, an intermediate regime may appear that is not captured by differential models~\cite{bosDynamicsSpectrallyTruncated2006}.

\section{Nonlocality of thermal damping}
\label{sec:damping}
Here we demonstrate analytically that, in the presence of a thermal bath, the damping rate of a localized perturbation to a thermal distribution in some cases diverges as the maximum wavenumber is taken to infinity.
The conditions for this divergence are derived for the three-wave and four-wave isotropic cases.
It will later be shown that this property can produce a nonlocal dependence in the depletion by thermal damping of a cascade.

\subsection{Three-wave interaction}
\label{sec:three-wave-nonlocality}
We consider a perturbation to thermal equilibrium which is localized in $k$-space,
and compute its linear damping rate.
We define $n_k=T/\omega(k)+\delta n_k$, and assume $\delta n_k \ll T/\omega(k)$ and that $\delta n_k$ is only nonzero in some localized region of $k$-space.
Plugging this into \cref{equ:WKE,equ:I-3} results in the linearized wave kinetic equation
\begin{widetext}
\begin{equation}
\partial_t \delta n_k= \pi \int\left[\left|V_{k 12}\right|^2 \delta f_{k 12} \delta\left(\boldsymbol{k}-\boldsymbol{k}_1-\boldsymbol{k}_2\right) \delta\left(\omega_k-\omega_1-\omega_2\right)-2\left|V_{1 k 2}\right|^2 \delta f_{1 k 2} \delta\left(\boldsymbol{k}_1-\boldsymbol{k}-\boldsymbol{k}_2\right) \delta\left(\omega_1-\omega_k-\omega_2\right)\right] d \boldsymbol{k}_1 d \boldsymbol{k}_2
\label{equ:WKE-linearized-three-wave}
\end{equation}
\end{widetext}
where
\begin{equation}
\begin{split}
\delta f_{k12}=&n_1 \delta n_2+\delta n_1 n_2\\
&-n_k\left(\delta n_1+\delta n_2\right)-\delta n_k\left(n_1+n_2\right).
\end{split}
\end{equation}
We are interested in the asymptotic behavior of this integral at $k\ll k_1$ or $k\ll k_2$, so we assume the following scalings:
\begin{equation}
    \lim_{k\to \infty}\omega(k)=A k^\alpha
    \label{equ:alpha}
\end{equation}
\begin{equation}
    \lim_{k/k_1\to 0}|V_{k12}|^2=V_0^2 k^{m_1}k_1^{2m-m_1}.
    \label{equ:three-wave-m}
\end{equation}
We will assume $\alpha\geq1$, allowing for the momentum and energy conservation constraints to be met.
Later, in the four-wave case, we will assume $\alpha<1$, precluding resonant three-wave coupling.

We note that $\delta n_k\sim 0$ for sufficiently large $k$, and that after accounting for  momentum conservation (enforced by a $\delta$-function), terms with $\delta n_1$ and $\delta n_2$ may be dropped.
We also note that due to energy conservation (also enforced by a $\delta$-function), the $V_{k12}$ process is not consistent with the limit in which $k$ is small, and the first term in the integral may be ignored.
The equation then simplifies in this limit to
\begin{equation}
\begin{split}
\partial_t \delta n_k=& -2\pi V_0^2 k^{m_1} \delta n_k\\
&\int k_1^{2m-m_1}(n_{1-k}-n_1) \delta\left(\omega_1-\omega_k-\omega_{1-k}\right) d \boldsymbol{k}_1.
\end{split}
\end{equation}
Applying $(n_{1-k}-n_1)\approx -\boldsymbol{k}\cdot\partial_{\boldsymbol{k}_1} n_1$, and integrating over the angular coordinates of $\boldsymbol{k_1}$, this yields
\begin{equation}
\partial_t \delta n_k\propto\frac{TV_0^2}{A^2}k^{m_1+\alpha-1}\int dk_1\,k_1^{2m-m_1+d-3\alpha}\ \delta n_k.
\end{equation}

This integral is dominated by a power of its upper integration bound for:
\begin{equation}
    2m-m_1+d-3\alpha>-1.
\end{equation}
In addition, when $2m-m_1+d-3\alpha>-1$, we find that the dominant component of the linear damping rate goes as
\begin{equation}
    \gamma \sim \frac{V_0^2}{A^2}Tk^{m_1+\alpha-1}k_m^{2m-m_1+d-3\alpha+1}
    \label{equ:three-wave-damping}
\end{equation}

\subsection{Four-wave interaction}
\label{sec:four-wave-nonlocality}

Analogously, linearizing \cref{equ:WKE_four-wave} yields
\begin{equation}
\begin{split}
     \partial_t\delta n_k=&\pi\int d\vec k_1\,d\vec k_2\,d\vec k_3\\
     &|V_{k123}|^2\delta f_{k123}\,\delta_{k1,23}\,\delta(\omega_k+\omega_1-\omega_2-\omega_3)
\end{split}
\end{equation}
where
\begin{equation}
\begin{split}
\delta f_{k123}=&\delta n_2\,n_3\left(n_k+n_1\right)+n_2n_3\,\delta n_k\\
&-\delta n_k\,n_1\left(n_2+n_3\right)-n_kn_1\,\delta n_2
\end{split}
\end{equation}

The relevant high wavenumber limit is $k,\ k_2 \ll k_1,\ k_3$, as the limit $k \ll k_1,\ k_2,\ k_3$ is precluded by energy conservation for waves with non-decay dispersion relations.
In this limit, we take
\begin{equation}
    \lim_{k,k_2 \ll k_1,k_3}|V_{k123}|=V_0 (k_1k_3)^{(m-m_1)/2}(kk_2)^{m_1/2}.
\end{equation}
A similar computation then yields that for $d+2m-2m_1-2\alpha>0$,
\begin{equation}
    \gamma_k=Gk^{d+2m_1-\alpha}\,k_m^{\,d+2m-2m_1-2\alpha}
    \label{equ:four-wave-damping}
\end{equation}
where $S_d$ is the area of the $d$-sphere and
\begin{equation}
    G=2\pi S_d^2\frac{T^2V_0^2}{\left(d+2m-2m_1-2\alpha\right)\alpha A^3}
\end{equation}

\section{Impact on warm cascades and non-universality}
\label{sec:intermediate}

While the above calculations concern the linear behavior of perturbations to a thermal distribution, we can expect that such dependence of thermal damping on the UV cutoff will carry over into the behavior of warm cascades as long as their short-wave asymptotic is thermal.
The central question is: \emph{Where does this thermal tail start?}
Naively, one might expect the transition between cascade and thermal regimes in a warm cascade to take place near the intersection point of these two stationary solutions.
This would produce a transition wavenumber $k_*$ dependent upon only flux and temperature.
However, \emph{the cascade may only exist where the rate of the interaction driving the cascade outpaces damping}.
As the damping rate increases with $k_m$, for sufficiently large $k_m$, the point at which the cascade rate and damping rate become equal (indicating a breakdown of the cascade) may be at lower wavenumber than the naive intersection point of the two asymptotic solutions.
Such circumstances suggest the presence of an intermediate region between these two regimes, whose bounds would then be dependent on the maximum wavenumber, breaking the assumption of universality.

\subsection{Three-wave interaction}
Assuming the collision integral to be dominated by local interactions, the characteristic time scale of the cascade may be estimated by
\begin{equation}
\frac{I_k[n]}{n_k}\sim \frac{V_0^2\lambda P^{1/2}}{A}k^{m-\alpha}.
\end{equation}
The ratio of the damping and cascade rates is then
\begin{equation}
\frac{\gamma n_k}{I_k[n]}\sim\frac{T}{A\lambda P^{1/2}}k_m^{2m-m_1+d-3\alpha+1}k^{m_1-m-1+2\alpha}.
\end{equation}
This ratio grows with $k$ for 
\begin{equation}
    m_1-m-1+2\alpha>0,
    \label{equ:three-wave-damping-ineq}
\end{equation}
reaching unity at
\begin{equation}
    k_c^{m_1-m-1+2\alpha}\sim\frac{A\lambda P^{1/2}}{T}k_m^{-2m+m_1-d+3\alpha-1}.
\end{equation}

Naively, the thermal and cascade solutions intersect at
\begin{equation}
     k_{\rm int}=\left(\frac{A\lambda P_0^{1/2}}{T}\right)^{1/(m+d-\alpha)}
     \label{equ:int_sound}
\end{equation}
such that
\begin{equation}
    k_c\sim\left(\frac{k_{\rm int}}{k_m}\right)^{\frac{2m-m_1+d-3\alpha+1}{m_1-m+2\alpha-1}}k_{\rm int}.
    \label{equ:k_c_three}
\end{equation}
Assuming the denominator to be positive (\cref{equ:three-wave-damping-ineq}), the damping outpaces the cascade before the cascade would intersect the thermal region if
\begin{equation}
    2m-m_1+d-3\alpha>-1.
    \label{equ:three-wave-balance-ineq}
\end{equation}
Any such system must possess a nontrivial intermediate range between the asymptotic cascade and thermal solutions.

\subsection{Four-wave interaction}
An analogous argument may be made for the energy cascade in the four-wave case.
Here the ratio of damping rate to cascade rate is
\begin{equation}
    \frac{\gamma_kn_k}{I_k[n]}\sim\frac{T^2}{A^2}V_0^{4/3}C^{2/3}\frac{k_m^{d+2m-2m_1-2\alpha}}{P^{2/3}}k^{d+2m_1-\frac{2}{3}m}
\end{equation}
which grows with $k$ for
\begin{equation}
    d+2m_1-\frac{2}{3}m>0.
    \label{equ:four-wave-damping-ineq}
\end{equation}
Naively, the thermal and cascade solutions intersect at $k_{\rm int}$ given by
\begin{equation}
    k_{\rm int}^{d+\frac23m-\alpha}=A\left(\frac{C}{V_0}\right)^{2/3}\frac{P_0^{1/3}}{T},
\end{equation}
and the ratio of rates reaches unity at
\begin{equation}
    k_c\sim\left(\frac{k_{\rm int}}{k_m}\right)^{\frac{d+2m-2m_1-2\alpha}{d+2m_1-\frac23m}}k_{\rm int}.
    \label{equ:k_c_four}
\end{equation}
Again, assuming the denominator in the exponent is positive (\cref{equ:four-wave-damping-ineq}), an intermediate range between the cascade and thermal ranges must exist if
\begin{equation}
    d+2m-2m_1-2\alpha>0.
    \label{equ:four-wave-balance-ineq}
\end{equation}

\section{Depletion of the cascade by damping}
\label{sec:depletion}
The presence of damping by nonlocal coupling to the thermal range at all points in the cascade results in a progressive loss of energy from the cascade and reduction in local flux.
In regions where the depletion is small, this may be modeled as a $k$-dependent $P(k)$ in the cascade solution \cref{equ:three-wave-cascade,equ:four-wave-cascade}.
When the depletion becomes larger, this approximation is no longer justified, but provides a limit to the distance to which the cascade can extend, which coincides with $k_c$ (\cref{equ:k_c_three,equ:k_c_four}) derived from time scales.

We first take \cref{equ:energy_continuity}, and consider the contribution from damping:
\begin{equation}
\begin{split}
    \partial_kP(k)=&-\partial_tE(k)\\
    =&(2k)^{d-1}\pi\omega_k\partial_tn_k\\
    =&-(2k)^{d-1}\pi\omega_k\gamma_kn_k.
\end{split}
\end{equation}
We then substitute the cascade solutions (\cref{equ:three-wave-cascade,equ:four-wave-cascade}) and $k$-dependent damping $\gamma_k$ (\cref{equ:three-wave-damping,equ:four-wave-damping}) for the three- and four-wave cases.
Following integration, we obtain an expression for the flux depleted by thermal damping.
Similar calculations with other forms of damping may be found in the literature, e.g. \cite{howesModelTurbulenceMagnetized2008,paoStructureTurbulentVelocity1965}.

\subsection{Three-wave interaction}
Substituting and integrating, we obtain
\begin{equation}
    \int_{P_0}^{P}P'^{-1/2}dP'=\int_{k_0}^{k}dk'\ -2^{d-1}\pi AG\lambda\,k'^{m_1-m+2\alpha-2},
\end{equation}
where $G$ is a pre-factor in $\gamma_k$.
This yields the solution
\begin{equation}
    P(k)=\left(P_0^{1/2}-\beta\left(k^{m_1-m+2\alpha-1}-k_0^{m_1-m+2\alpha-1}\right)\right)^2,
\end{equation}
where
\begin{equation}
    \beta=\frac{2^{d-2}\pi AG\lambda}{m_1-m+2\alpha-1}.
\end{equation}

$P(k)$ starts at a value $P_0$ at the scale $k_0$ where energy is injected, and decreases with increasing $k$ as damping acts as a distributed sink.
At some $k_d$, the flux is completely depleted,
\begin{equation}
    k_d^{m_1-m+2\alpha-1}=\frac{P_0^{1/2}}{\beta}+k_0^{m_1-m+2\alpha-1}\approx\frac{P_0^{1/2}}{\beta},
\end{equation}
and we may write the scale-dependent flux as
\begin{equation}
    P(k)\approx P_0\left(1-\left(\frac{k}{k_d}\right)^{m_1-m+2\alpha-1}\right)^{2}
\end{equation}
We see that this solution describes a depleting cascade when $m_1-m+2\alpha-1>0$, the condition we derived earlier for the increasing dominance of damping over the cascade with $k$ (\cref{equ:three-wave-damping-ineq}).
Further, a rearrangement of coefficients and definitions yields
\begin{equation}
    k_d^{\,m_1-m+2\alpha-1}\sim\left(\frac{k_{\rm int}}{k_m}\right)^{2m-m_1+d-3\alpha+1}k_{\rm int}^{\,m_1-m+2\alpha-1}.
\end{equation}
Comparing to \cref{equ:k_c_three}, we see that $k_c\sim k_d$, and $k_d<k_{\rm int}$ for the condition given by \cref{equ:three-wave-balance-ineq}.
The cascade depletes at \emph{lower} $k$, as the maximum wavenumber is sent to \emph{higher} $k$.

The location $k_*$ of the intersection of the depleted cascade with the thermal distribution is then given by the equation
\begin{equation}
\left(\frac{k_*}{k_{\rm int}}\right)^{m+d-\alpha}=1-\left(\frac{k_*}{k_d}\right)^{m_1-m+2\alpha-1}.
\end{equation}
$m+d-\alpha>0$ if \cref{equ:three-wave-balance-ineq,equ:three-wave-damping-ineq} are met.
Since the depleted cascade vanishes at $k_d$, the solution must have $k_*<k_{\rm int}$.
The transition $k_*$ between the cascade and thermal solutions is at a lower wavenumber than naive intersection would suggest, and depends on the small scale physics.

\subsection{Four-wave interaction}
Integrating in similar fashion leads to
\begin{equation}
    P(k)=\left(P_0^{2/3}-\beta\left(k^{d+2m_1-\frac23m}-k_0^{d+2m_1-\frac23m}\right)\right)^{3/2},
\end{equation}
where
\begin{equation}
\beta=\frac{2^{d}\pi AG}{3(d+2m_1-\frac23m)}\left(\frac{C}{V_0}\right)^{2/3}.
\end{equation}
Again, we get a depletion scale,
\begin{equation}
    k_d^{d+2m_1-\frac23m}=\frac{P_0^{2/3}}{\beta}+k_0^{d+2m_1-\frac23m}\approx\frac{P_0^{2/3}}{\beta},
\end{equation}
and a reduced form for the scale-dependent flux,
\begin{equation}
    P\approx P_0\left(1-\left(\frac{k}{k_d}\right)^{d+2m_1-\frac23m}\right)^{3/2}.
\end{equation}
Again, the exponent matches that derived earlier (\cref{equ:four-wave-damping-ineq}), and a short rearrangement yields that $k_d\sim k_c$ (compare to \cref{equ:k_c_four}) and the inherited condition on the exponent (\cref{equ:four-wave-balance-ineq}):
\begin{equation}
    k_d^{d+2m_1-\frac23m}=\frac{a}{C^2}\left(\frac{k_{\rm int}}{k_m}\right)^{d+2m-2m_1-2\alpha}k_{\rm int}^{\,d-\frac23m+2m_1}.
\end{equation}
(Since for the four-wave case the angular part of the linearized collision integral simplifies, we have given $\gamma_k$ in a closed form \cref{equ:four-wave-damping}, which results in a dimensionless product of coefficients $a$.)

$k_*$ is determined by the equation
\begin{equation}
\left(\frac{k_*}{k_{\rm int}}\right)^{d+\frac23m-\alpha}=\left(1-\left(\frac{k_*}{k_d}\right)^{d+2m_1-\frac23m}\right)^{1/2}.
\end{equation}
$d+\frac23m-\alpha>0$ if \cref{equ:four-wave-balance-ineq,equ:four-wave-damping-ineq} are met.
Once again, this implies $k_*<k_{\rm int}$, and the transition is shifted to a lower $k$ in a manner dependent on the UV cutoff.

\section{Acoustic turbulence}
To make the above discussion more concrete, and to enable numerical testing of our model, we now give the same treatment of a particular case with three-wave coupling, longitudinal acoustic modes in an isotropic three-dimensional medium at temperatures much larger than the Debye temperature~\cite{zakharovKolmogorovZakharovSpectraTurbulence2025}.
In this case, the frequency is
\begin{equation}
    \omega_k=ck,
    \label{equ:freq_sound}
\end{equation}
and the three-wave coupling is
\begin{equation}
    V_{k12}=\sqrt{bkk_1k_2},
    \label{equ:V_sound}
\end{equation}
where $b=c/\rho$ for sound speed $c$ and density  $\rho$.
Integrating over wave vector angle and retaining magnitude $k$, the collision integral may be written
\begin{widetext}
\begin{equation}
I_k[n]=2 \pi^2 \frac{b}{c}\left(\int_0^k k_1^2\left(k-k_1\right)^2 f_{k\ 1\ k-1} d k_1-2 \int_k^{k_m} k_1^2\left(k_1-k\right)^2 f_{1\ k\ 1-k} d k_1\right).
\label{equ:WKE-acoustic}
\end{equation}
\end{widetext}
The stationary solutions to this equation are equilibrium $n_k=T/ck$ for temperature $T$ and sound speed $c$, and cascade $n_k=\lambda P^{1/2}k^{-9/2}$ for energy flux $P$, where $\lambda=Cb^{-1/2}$ with a constant $C=\left(\frac{256\pi^3}{3}\left(\ln16+\pi-1\right)\right)^{-1/2}$~\cite{zakharovSpectrumAcousticTurbulence1970,zakharovKolmogorovZakharovSpectraTurbulence2025,kochurinDirectNumericalSimulation2022,kochurinAcousticTurbulenceZakharov2025}.

\subsection{Analytic model}
From \cref{equ:alpha,equ:three-wave-m,equ:freq_sound,equ:V_sound}, and our choice of dimension, we extract the exponents:
\begin{equation}
\begin{split}
d=&3,\\
\alpha=&1,\\
m=&\frac32,\\
m_1=&1.
\end{split}
\end{equation}
This gives for \cref{equ:three-wave-damping-ineq,equ:three-wave-balance-ineq}
\begin{align}
    m_1-m-1+2\alpha=1/2>&0,\\
    2m-m_1+d-3\alpha=2>&-1.
\end{align}

Per \cref{sec:damping}, we calculate the damping rate to be
\begin{equation}
    \gamma=\frac{4}{3}\pi^2\frac{b}{c^2}Tk_m^3k\left(1-\frac{3}{2}\frac{k}{k_m}+\frac{3}{4}\frac{k^3}{k_m^3}\right).
\end{equation}
We show here subdominant contributions dropped in \cref{sec:damping} for completeness, and will not utilize them in later calculations.
The leading term is consistent up to a constant with the damping rate found in the literature \cite[Ch.~5]{gurevichSound1986}.
Applying the argument of \cref{sec:depletion}, we find
\begin{equation}
\begin{split}
    P(k)=&\left(P_0^{1/2}-\beta\left(k^{1/2}-k_0^{1/2}\right)\right)^2,\\
    \beta=&\frac{16}{3}\pi^3\frac{b}{c}T\lambda k_m^3,
\end{split}
\label{equ:P_sound}
\end{equation}
This gives
\begin{equation}
    P(k)\approx P_0\left(1-\left(\frac{k}{k_d}\right)^{1/2}\right)^2,
\end{equation}
with
\begin{align}
    k_d=&\left(M\frac{k_{\text{int}}}{k_m}\right)^6k_{\text{int}},\label{equ:k_d_sound}\\
    M=&\left(\frac{16\pi^3}{3}C^2\right)^{-1/3}\approx4.28,
\end{align}
and
\begin{equation}
    \left(\frac{k_*}{k_{\text{int}}}\right)^{7/2}=1-\left(\frac{k_*}{k_d}\right)^{1/2}
    \label{equ:unnorm_sound}
\end{equation}
For the purposes of later discussion, we will normalize with respect to $k_{\rm int}$:
\begin{equation}
    \kappa=\frac{k}{k_{\rm int}}.
\end{equation}
Rearranging \cref{equ:unnorm_sound} and using \cref{equ:k_d_sound} yields an implicit equation for $\kappa_*$ in terms of $\kappa_m$, which is the prediction we will evaluate numerically:
\begin{equation}
    \kappa_*^{-1/2}-\kappa_*^3=\left(\frac{\kappa_m}{M}\right)^3
    \label{equ:norm_sound}
\end{equation}

\subsection{Numerical approach}
In order to study a non-equilibrium state, we must add sources and sinks.
We search for stationary solutions of
\begin{equation}
    \partial_t n_k=I_k[n]+\frac{P_0}{4\pi\omega_{k_0} k_0^2}\delta(k-k_0)+\sigma (T_0-\omega_k n_k).
    \label{equ:EoM_sim}
\end{equation}
Energy is injected isotropically at a minimum wavenumber $k_0$ at rate $P_0$, and is removed by a thermostat with strength parameterized by $\sigma$ which attempts to return the system to thermal equilibrium at temperature $T_0$.
This system is expected to relax to a stationary solution with a direct energy cascade from $k_0$ to larger wavenumber, which transitions to a thermal range at high wavenumber.
The degree of deviation of the thermal range from temperature $T_0$ is determined by the parameter $\sigma$.

\subsubsection{Integration algorithm}
\Cref{equ:EoM_sim} was solved on a grid between $k_0=1$ and some integer $k_m$.
The system under consideration is an integral equation covering a wide range of scales, and is therefore very stiff.
Instead of directly integrating \cref{equ:EoM_sim}, it was more effective to relax from some initial condition to a stationary equilibrium along another trajectory.
This was performed using a Rosenbrock-type method with an approximate Jacobian~\cite{steihaugAttemptAvoidExact,rosenbrockGeneralImplicitProcesses1963,hairerRosenbrockTypeMethods1996}.
This produces a stationary solution to the original equation of motion with more efficient steps.

\subsubsection{Fixing variables and initial conditions}
We are interested in demonstrating the dependence of the transition point on the maximum wavenumber $k_m$ with other properties fixed.
$P_0$ was fixed by \cref{equ:int_sound} to produce a particular naive $k_{\rm int}$ against a thermal range with $T_0$, with the deviation of the true transition point $k_*$ from $k_{\rm int}$ measured as $k_m$ was varied between runs.

We would like to measure the depletion of the cascade against a constant thermal background near $T_0$.
However, the thermostat cannot remove energy if $n_k$ is exactly a thermal distribution at $T_0$, and the stationary state will necessarily deviate from this.
In order to maintain control over this deviation, we choose a target value of $T$ (measured as described in \cref{sec:measurements}) near $T_0$, and tune $\sigma$ in each run to maintain it.
Each run was performed first with an adaptive loop to set $\sigma$ and allowed to converge, then run with frozen $\sigma$ at the converged value for at least the same number of steps and compared to confirm that the stationary solution does not change.
All runs shifted by no more than $0.5\%$ at any wavenumbers.

Runs were performed with thermostat $T_0=1$, target $T=1.01$ (setting $\sigma$), $k_{\rm int}=1000$ (fixing by $P_0$ given $T$), and $k_m$ varied from $2750$ to $8650$.
The initial conditions were the sum of a depleted cascade as given by \cref{equ:P_sound} and a thermal distribution at $T$: $n_k=\lambda P(k)^{1/2}k^{-9/2}+T/ck$.
As a control, additional runs were performed for $k_m=3000,\ 4250,\ 5500,\ 6750,\ 8000$ starting from a sum of an (un-depleted) cascade and a thermal distribution: $n_k=\lambda P_0^{1/2}k^{-9/2}+T/ck$, and were found to converge to within $3\%$ of the primary runs at all wavenumbers.

\subsubsection{Compensating for the thermostat}
The thermostat is present across the domain, such that it depletes the cascade alongside damping by the thermal region.
This contaminates the signal we are attempting to measure.
As such, we derive an effective damping and resulting adjusted scaling.

For $n_k\gg \frac{T}{\omega_k}$ (in the cascade region), the thermal damping and thermostat terms have the same form,
\begin{equation}
    \partial^{\rm(dissipation)}_tn_k=-\gamma n_k-c\sigma k n_k.
\end{equation}
We therefore incorporate the thermostat into an effective damping.
\begin{equation}
    \gamma_{\text{eff}}=\gamma+c\sigma k
\end{equation}
Going through the same steps as before, we arrive at an analogue of \cref{equ:norm_sound},
\begin{equation}
    \kappa_*^{-1/2}-\kappa_*^3=\left(\frac{\kappa^{\rm eff}_m}{M}\right)^3
    \label{equ:renorm_sound},
\end{equation}
where
\begin{equation}
    \kappa^{\rm eff}_m=\kappa_m\left(1+\frac{4\pi c^2\lambda\sigma}{\beta}\right)^{1/3}.
\end{equation}

We will use a figure of merit between $0$ and $1$ indicating the extent to which the thermostat is contributing to damping of the cascade as a means of comparison of runs:
\begin{equation}
    R=\frac{\sigma c k_*}{\gamma_{k_*}+\sigma c k_*}
    \label{equ:R}
\end{equation}

\subsubsection{Measurements}
\label{sec:measurements}
\paragraph\ Transition wavenumber ($k_*$):
The transition wavenumber was calculated from the local scaling exponent of $n_k$:
\begin{equation}
    s(k)=-\frac{d\ln n_k}{d\ln k}.
    \label{equ:exp}
\end{equation}
The depleted cascade is expected to have $s(k)$ steadily increase from $9/2$, before transitioning to a constant $1$ in the thermal region.
Assuming smooth $n_k$, the transition region must therefore contain an inflection point $s''(k)=0$, which was used as a proxy for $k_*$.
\paragraph\ Temperature ($T$):
The temperature was measured as the median value of $ck n_k$ over the range between $3k_*/2$ and $k_m$ for the purpose of fixing $\sigma$, and by a fit to $T/ck$ when calculating $R$.

\subsection{Results of numerics}
\begin{figure*}
    \centering
    \includegraphics[width=\textwidth]{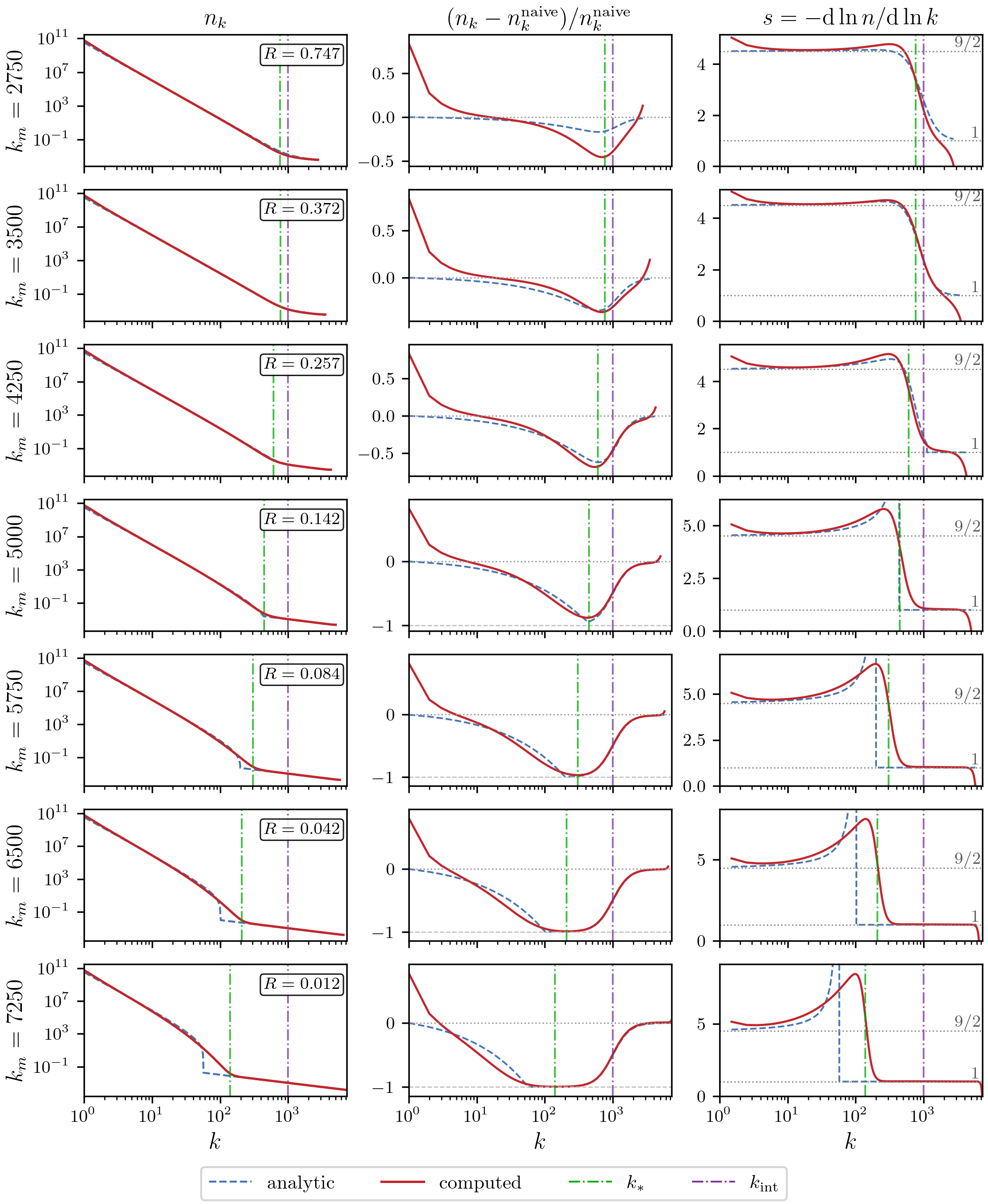}
    \caption{
    Spectrum $n_k$, difference of the spectrum from a naive sum $n^{\rm naive}_k=\lambda P_0^{1/2}k^{-9/2}+T/ck$, and local exponent (\cref{equ:exp}) for the stationary numerical solutions at evenly spaced $k_m$.
    The sum of the depleted cascade and a thermal distribution $n_k=\lambda P(k)^{1/2}k^{-9/2}+T/ck$ (using \cref{equ:P_sound}) are shown for reference.
    The thermostat contamination parameter $R$ (\cref{equ:R}) is indicated in the left column.
    }
    \label{fig:n_k}
\end{figure*}

\begin{figure*}
    \centering
    \includegraphics[width=\textwidth]{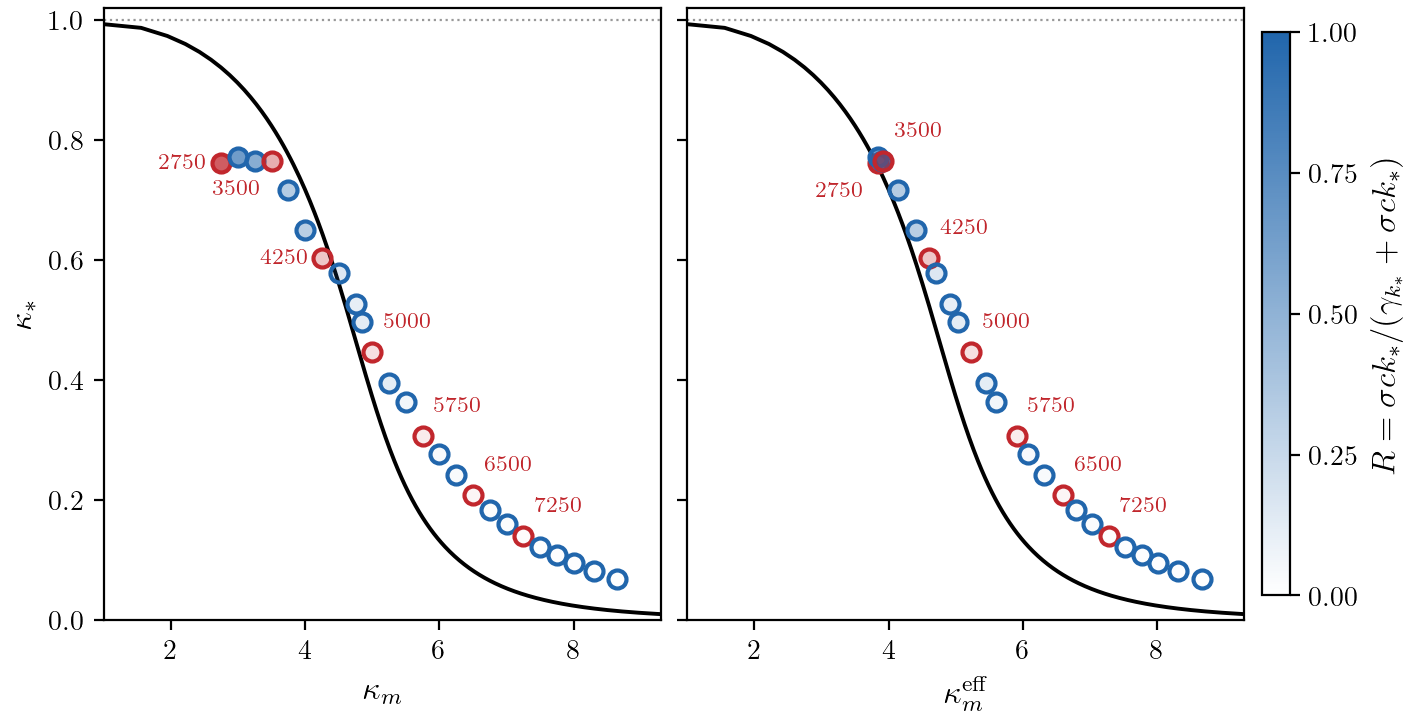}
    \caption{
    Normalized transition wavenumber as a function of uncompensated (left) and compensated (right) normalized maximum wavenumber.
    Opacity is used to indicate thermostat contamination parameter $R$.
    Red markers indicate the runs shown in figure \cref{fig:n_k} with indicated $k_m$.
    The curves defined by \cref{equ:norm_sound,equ:renorm_sound} are shown in black.
    }
    \label{fig:k}
\end{figure*}

The primary qualitative result is visible in \cref{fig:n_k}.
The transition between the cascade and thermal regions, visible as a knee in the spectrum and a sharp drop in the exponent, moves to lower wavenumber as the maximum wavenumber is increased.
Further, it can be seen that, at least for a moderate range of $k_m$, the depleted cascade model of $n_k$ captures some of the qualitative features of the numerical results.
In particular, a transition region with increased slope is present at a comparable location to the analytic result.
While we have captured the presence of such a transition region, our approximation extrapolates from the impact on the cascade, and cannot describe the transition region itself.
Our analytic model for the dependence of transition wavenumber on maximum wavenumber, given by \cref{equ:norm_sound,equ:renorm_sound}, is tested in \cref{fig:k}.
Qualitative agreement can be seen, particularly for $\kappa_m^{\rm eff}\approx3-5$.

The increasing discrepancies at higher and lower $k_m$ seen in both figures are an inherent product of the numerics.
At large $k_m$, the depletion of the cascade moves to sufficiently low wavenumbers that there is no longer a true inertial range with a cascade.
This is visible in \cref{fig:n_k} in the lack of an extended interval with $s(k)$ near $9/2$ for large $k_m$.
At small $k_m$, there is no longer a substantial thermal range over which the thermostat can absorb the injected energy, and the power budget is balanced by transfer from the cascade to the thermostat, counteracting the reduced damping by the thermal region.
Indeed, in \cref{fig:k}, it can be seen that the uncompensated value of $k_*$ is roughly constant as $k_m$ is varied for runs where the contamination parameter $R$ is large.
Lower values of $\kappa_m^{\rm eff}$ cannot be probed without changing other variables, as the compensated values for reduced $k_m$ collapse to a point.
The presence of differing processes at low, moderate, and large $k_m$ relative to a characteristic scale is qualitatively in line with \cref{equ:norm_sound}, whose solution is a sigmoid (plotted in \cref{fig:k}).

\section{Discussion}

We demonstrate by scaling arguments and a numerical counterexample that, contrary to an assumption commonly made in the literature,  stationary solutions to the wave kinetic equation describing a cascade at nonzero temperature are not necessarily parameterized solely by flux and temperature.
Rather, as we demonstrate numerically for the case of sound waves in three dimensions, the form of the solution may depend on the small scale physics of the system, namely the maximum wavenumber.
This is due to the nonlocality of thermal damping.

Interactions in KZ solutions of the wave kinetic equation are local in the sense that the collision integral converges as its limits are taken to zero and infinite wavenumber.
This is due to a felicitous cancellation of different regions of integration~\cite{zakharovKolmogorovZakharovSpectraTurbulence2025}.
Meanwhile, in the thermal solution, the integrand vanishes identically across the domain due to detailed balance.
Therefore there is no reason to assume that the behavior of perturbations to this distribution behave locally.
As shown in \cref{sec:damping}, there are indeed cases in which the evolution of a localized perturbation depends on all scales.
As a result, cascade and thermal portions cannot be treated as independent.

Further insight into this result can be gained by considering the neutrally stable infinitesimal modifications to one-parameter spectra (equilibrium or cascade). This gives an additional illustration on how misleading are the results obtained from differential approximations of the collision integral,
In differential models, the equilibrium and cascade asymptotics have the same number of neutral modes.
This is surprisingly not the case for general kinetic equations with integral collision operators.
While thermal spectra admit stationary (neutral, marginal) corrections that carry conserved fluxes, turbulent spectra do not admit stationary corrections that correspond to equilibrium parameters.
For the three-wave kinetic equation, thermal equipartition admits a stationary small correction with a constant energy flux, while the turbulent cascade does not admit any small stationary isotropic corrections at all (except trivial flux change)~\cite{zakharovKolmogorovZakharovSpectraTurbulence2025}. 
Universality exists in equilibrium and near it. The universality is lost far from thermal equilibrium. The ultimate turbulent state seems universal, but however small deviations from it are not.
Deviations from the cascade solution due to nonzero temperature instead must have a non-universal form, the outer edge of the intermediate region, for which we propose an approximate expression in \cref{sec:depletion}.
This approximation is seen in the numerics to reflect qualitative features of the transition, but does not describe its location and shape precisely.

\bibliography{references}
\bibliographystyle{abbrv}

\end{document}